\documentclass[12pt]{article}
\usepackage{amsmath}
\usepackage{amsfonts}
\usepackage{amssymb}
\usepackage{graphicx}

\begin{document}

\author{A. Wehner\thanks{Electronic-mail: sllg7@cc.usu.edu}\\Department of Physics, Utah State University, Logan, Utah 84322}
\title{Symmetric Spaces with Conformal Symmetry}
\maketitle
\begin{abstract}
We consider an involutive automorphism of the conformal algebra and the
resulting symmetric space. We display a new action of the conformal group
which gives rise to this space. The space has an intrinsic symplectic
structure, a group-invariant metric and connection, and serves as the model
space for a new conformal gauge theory.
\end{abstract}

\pagebreak 

\section{Introduction}

Symmetric spaces are the most widely studied class of homogeneous spaces. They
form a subclass of the reductive homogeneous spaces, which can essentially be
characterized by the fact that they admit a unique torsion-free
group-invariant connection. For symmetric spaces the curvature is covariantly
constant with respect to this connection. Many essential results and extensive
bibliographies on symmetric spaces can be found, for example, in the standard
texts by Kobayashi and Nomizu \cite{Kobayashi} and Helgason \cite{Helgason}.

The physicists' definition of a (maximally) symmetric space as a metric space
which admits a maximum number of Killing vectors is considerably more
restrictive. A thorough treatment of symmetric spaces in general relativity,
where a pseudo-Riemannian metric and the metric-compatible connection are
assumed, can be found in Weinberg \cite{Weinberg}. The most important
symmetric space in gravitational theories is Minkowski space, which is based
on the symmetric pair (Poincar\'{e} group, Lorentz group). Since the
Poincar\'{e} group is not semi-simple, it does not admit an intrinsic
group-invariant metric, which would project in the canonical fashion to the
Minkowski metric $\eta_{ab}=diag(-1,1\ldots1)$.

The indisputable importance of conformal symmetry in physics has led us to
describe a particular class of symmetric spaces which are built from the
conformal group and admit a group-invariant metric as well as rich additional
structures. There are compelling physical arguments that the conformal group,
not the Poincar\'{e} group, should be the fundamental symmetry group of
spacetime, not the least of which is the fact that it includes scale
transformations (expressing our freedom to choose units arbitrarily) as a
fundamental symmetry of nature. In addition, the conformal group is
mathematically more natural than the Poincar\'{e} group as it is a
\textit{simple} Lie group, which admits a group-invariant indefinite
Riemannian metric.

In contrast to the standard homogeneous space built from the conformal group
(briefly described in Sec. 2), our construction takes the symmetric structure
of the conformal Lie algebra into account (Sec. 3). The result is an
even-dimensional symmetric space with local Lorentz and scaling symmetry. It
has a conformally invariant connection and a natural symplectic structure. It
can serve as the model geometry for a generalized space (in the sense of
Cartan) called \textit{biconformal space}, which was first discussed by
Wheeler \cite{Wheeler1}. We present a nonlinear action of the conformal group
on this space, which treats translations and special conformal transformations symmetrically.

\section{Conformal Group and Homogeneous Spaces}

The conformal group $C(n)$\ is the group of transformations preserving angles
between vectors when acting on $n$-dimensional ($n>1$) compactified Minkowski
space $(\mathcal{M}^{n},\eta_{\mu\nu})$ with metric $\eta_{ab}=diag(-1,1\ldots
1)$, $a,b,\ldots=1\ldots n$. Let $\{x^{\mu},\mu=1\ldots n\}$ be local
coordinates for $\mathcal{M}^{n}$. For $n>2$, the case we are considering from
now on, the conformal group possesses the following well-known nonlinear
action on $\mathcal{M}^{n}$:
\begin{align}
M_{\nu}^{\mu}  & =-\Delta_{\beta\nu}^{\mu\alpha}x^{\beta}\partial_{\alpha
}\label{VF1}\\
D  & =x^{\mu}\partial_{\mu}\\
P_{\mu}  & =\partial_{\mu}\\
K^{\mu}  & =\left(  x^{\mu}x^{\nu}-\tfrac{1}{2}x^{2}\eta^{\mu\nu}\right)
\partial_{\nu}\label{VF4}%
\end{align}
where $\partial_{\mu}\equiv\tfrac{\partial}{\partial x^{\mu}}$ and we defined
the antisymmetrization operator
\[
\Delta_{\mu\nu}^{\alpha\beta}\equiv\tfrac{1}{2}(\delta_{\mu}^{\alpha}%
\delta_{\nu}^{\beta}-\eta^{\alpha\beta}\eta_{\mu\nu}).
\]
These $\tfrac{1}{2}(n+1)(n+2)$ vector fields are the solutions to the
conformal Killing equation in $\mathcal{M}^{n}$, i.e., the vector fields that
preserve $\eta_{\mu\nu}$ up to an overall function. They generate Lorentz
transformations, scalings (dilatations), translations, and special conformal
transformations, respectively. The latter are simply translations of the
inverse coordinate $z^{\mu}\equiv-x^{\mu}/x^{2}$ (hence the need for
$\mathcal{M}^{n}$ to be compact). The commutation relations of the vector
fields (\ref{VF1})-(\ref{VF4}) are%
\[%
\begin{array}
[c]{ccc}%
\lbrack M_{\beta}^{\alpha},M_{\nu}^{\mu}]\;=\Delta_{\nu\beta}^{\alpha\gamma
}M_{\gamma}^{\mu}-\Delta_{\gamma\beta}^{\alpha\mu}M_{\nu}^{\gamma} &  &
[D,P_{\alpha}]\;=\;-P_{\alpha}\\
\lbrack M_{\beta}^{\alpha},P_{\mu}]\;=\Delta_{\mu\beta}^{\alpha\gamma
}P_{\gamma} &  & [D,K^{\alpha}]\;=\;K^{\alpha}\\
\lbrack M_{\beta}^{\alpha},K^{\mu}]\;=-\Delta_{\gamma\beta}^{\alpha\mu
}K^{\gamma} &  & [P_{\alpha},K^{\beta}]\;=D\delta_{\alpha}^{\beta}-2M_{\alpha
}^{\beta}%
\end{array}
\]
All other commutators vanish. These commutators are the same as those of the
infinitesimal generators of the pseudo-orthogonal group $O(n,2)$ in an
appropriate basis, i.e., $C(n)$\ and $O(n,2)$ are locally isomorphic. We note
that our mixed index positioning, which originates from the $O(n,2)$ matrix
notation, does \textit{not} imply any use of the metric, but rather indicates
the conformal weight of the operators as measured by the dilatation generator
$D$. Every upper index contributes $+1$ to the conformal weight of the
operator, while every lower index contributes $-1$. Thus, $K_{a}$ and $P^{a}$
have conformal weight $+1$ and $-1$, respectively, while $D$ and $M_{\beta
}^{a}$ are weightless.

The isotropy subgroup of the infinitesimal action (\ref{VF1})-(\ref{VF4}) at
the origin is generated by those vector fields that vanish at the origin,
namely $\{M_{\nu}^{\mu},D,K^{\mu}\}$. It is the Poincar\'{e} group extended by
scalings, which is commonly called the (inhomogeneous) similarity group,
$IS(n)$. Since the action is transitive on $\mathcal{M}^{n}$, the space
$\mathcal{M}^{n}$ is isomorphic to the $n$-dimensional homogeneous space
$C(n)/IS(n)$. The space $\mathcal{M}^{n}$ has long been known as
\textit{conformal (or M\"{o}bius) space}; it is topologically equivalent to a
(pseudo-)sphere. 

Curved generalizations of this space were first considered in 1923 by Elie
Cartan \cite{Cartan}. In modern language, Cartan's generalization of conformal
space amounts to a principal bundle $P\rightarrow M^{n}$ of dimension
\[
\dim P=\dim C(n)=\tfrac{1}{2}(n+1)(n+2),
\]
with fiber $IS(n)$ and an $n$-dimensional curved base space $M^{n}$ equipped
with a conformal structure, i.e., an equivalence class of conformally related
metrics. The bundle space $P$ may be interpreted as the first differential
prolongation (in the sense of Kobayashi \cite{Kobayashi2}) of the conformal
structure. The tangent space at any point of $P$ is isomorphic to the
conformal algebra, with the isomorphism given by the \textit{conformal
connection}, historically one of the first examples of what is today known as
a Cartan connection. Later on, spaces with conformal connections were
considered by many other authors \cite{Akivis}. 

Physicists have extensively used generalized conformal spaces in the 1970s as
background geometries for conformal gauge theories \cite{Ferber}. Such
theories were all shown to reduce to gauge theories formulated on Weyl
geometries, in the sense that the connection forms (gauge fields)
corresponding to the special conformal generators are algebraically removable
in any field theory. In fact, it has been claimed on geometrical and physical
grounds that the gauge transformations generated by special conformal
transformations are redundant altogether, as their inclusion seems to amount
either to gauging the same symmetry twice or to associating them with an
unknown external symmetry of nature \cite{Mansouri}. We do not wish to comment
on this claim, but remark that even though general Weyl geometries incorporate
local scale-invariance as an important physical principle, they do not give
rise to general relativity in arbitrary dimensions without the use of
additional structures such as compensating fields. Even more problematic is
the fact that they predict unphysical size changes.

\section{Symmetric Spaces}

Recall that an involutive automorphism (of order $2$) of a Lie group $G$ is a
Lie group automorphism $\sigma:G\rightarrow G$ such that\ $\sigma^{2}=1$,
$\sigma\neq1$. If $H$ is a Lie subgroup of $G$ with involutive automorphism
$\sigma$ such that $H$ is fixed by $\sigma$, then the coset space $G/H$ is
called a \textit{symmetric space}. Let $\frak{g}$ and $\frak{h}$ be the Lie
algebras of $G$ and $H$ and the Lie algebra automorphism $\sigma_{\frak{g}%
}:\frak{g}\rightarrow\frak{g}$ the one induced by $\sigma$. We have the
decomposition $\frak{g}=\frak{h}\oplus\frak{m}$ (direct sum), where
$\frak{h}\equiv\{X\in\frak{g}:\sigma(X)=X\}$ forms a Lie subalgebra and
$\frak{m}\equiv\{X\in\frak{g}:\sigma(X)=-X\}$ is the complementary space. As a
consequence, $[\frak{h},\frak{m}]\subset\frak{m}$ and $[\frak{m}%
,\frak{m}]\subset\frak{h}$.\ The pair $(\frak{g},\frak{h})$ is are known as a
\textit{symmetric pair}. The vector space spanned by $\frak{m}$ can be
identified in a natural way with both the homogeneous space $\frak{g}%
/\frak{h}$ and the tangent space at the point $H$ of $G/H$.

Homogeneous conformal spaces have a Lie algebra decomposition such that
$\frak{h}=\{M_{\nu}^{\mu},D,K^{\mu}\}$ and $\frak{m}=\{P_{\mu}\}$. Because
$[P_{\alpha},K^{\beta}]\;=D\delta_{\alpha}^{\beta}-2M_{\alpha}^{\beta}$,
$[\frak{h},\frak{m}]\nsubseteqq\frak{m}$, so these spaces are not symmetric.
The reason is that homogeneous conformal spaces, while retaining the largest
possible continuous symmetry on the fibers, do not take the discrete
symmetries of the conformal algebra into account. The conformal algebra admits
the following two involutive automorphisms:
\begin{equation}
\sigma_{1}:\left.
\begin{array}
[c]{l}%
P_{\alpha}\rightarrow-P_{\alpha}\\
K^{\alpha}\rightarrow-K^{\alpha}\\
D\rightarrow D\\
M_{\beta}^{\alpha}\rightarrow M_{\beta}^{\alpha}.
\end{array}
\right.  \qquad\sigma_{2}:\left.
\begin{array}
[c]{l}%
P_{\alpha}\rightarrow-\eta_{\alpha\beta}K^{\beta}\\
K^{\alpha}\rightarrow-\eta^{\alpha\beta}P_{\beta}\\
D\rightarrow-D\\
M_{\beta}^{\alpha}\rightarrow M_{\beta}^{\alpha}.
\end{array}
\right.  \label{discreete}%
\end{equation}
Both transformations leave the Lorentz sector of the conformal algebra
invariant, but only the first incorporates in addition
scale-invariance.\footnote{A third involutive automorphisms is given by the
composition of $\sigma_{1}$ and $\sigma_{2}$. Since the conformal group is
non-compact and semisimple, its Lie algebra admits in addition a
\textit{Cartan involution}, i.e. an involutive automorphism whose fixed set is
the Lie algebra of the maximal compact subgroup of $C(n)$. Symmetric spaces
defined by Cartan involutions are known to be homeomorphic to Euclidean spaces
\cite{Helgason}.} Thus, we shall be interested in symmetric spaces based on
$\sigma_{1}$. In this case, the Lie subgroup $H$ is the (non-compact)
similarity, homothety, or conformal orthogonal group $S(n)$, i.e., the direct
product of the Lorentz group and the scaling group, $S(n)=O(n-1,1)\otimes
\mathbf{R}^{+}$. The resulting symmetric space,
\[
B(n):=C(n)/S(n),
\]
is of dimension $2n$. The conformal group is simple, so $S(n)$ cannot contain
a non-trivial normal subgroup of $C(n)$. Thus, $C(n)$ acts not only
transitively, but also effectively on $B(n)$. The tangent space $T_{e}(B(n))$
at the identity is isomorphic to the vector space $\frak{m}$ spanned by
$P_{\alpha}$ and $K^{\alpha}$. The space $B(n)$ has an intrinsic metric which
follows from the Killing form of the conformal algebra. The Killing form of
the conformal group in the chosen basis is
\[
K_{conf}=2n\left(
\begin{array}
[c]{cccc}%
\tfrac{1}{2}\Delta_{\mu\nu}^{\alpha\beta} & 0 & 0 & 0\\
0 & 0 & \delta_{\alpha}^{\beta} & 0\\
0 & \delta_{\beta}^{\alpha} & 0 & 0\\
0 & 0 & 0 & 1
\end{array}
\right)
\]
Since the conformal group is semi-simple and non-compact, its Killing form is
non-degenerate and indefinite and hence gives rise to a conformally invariant
indefinite Riemannian metric on $C(n)$. The vector spaces $\frak{h}$ and
$\frak{m}$ are mutually orthogonal with respect to $K_{conf}$. We restrict
this quadratic form to the space $B(n)$ and normalize it:
\begin{equation}
K_{\alpha\beta}=\tfrac{1}{2}\left(
\begin{array}
[c]{cc}%
0 & \delta_{\alpha}^{\beta}\\
\delta_{\beta}^{\alpha} & 0
\end{array}
\right)  \label{Killing}%
\end{equation}
This restricted metric is still nondegenerate. Given a vector $v=v^{\alpha
}P_{\alpha}+v_{\alpha}K^{\alpha}$, we have $K(v,v)=v^{\alpha}v_{\alpha}$ which
shows that $K$ is indefinite on $B(n)$.

We now give an action of $C(n)$ on the space $B(n)$. Let $\{x^{\mu},y_{\mu}\}$
be local coordinates for $B(n)$, and let $(\eta_{\mu\nu},-\eta^{\mu\nu})$,
which is the diagonalized form of the restricted Killing metric (\ref{Killing}%
), be given on $B(n)$. Then an infinitesimal action is given by the vector
fields
\begin{align*}
M_{\nu}^{\mu}  & =-\Delta_{\beta\nu}^{\mu\alpha}\left(  x^{\beta}%
\partial_{\alpha}-y_{\alpha}\partial^{\beta}\right)  \\
D  & =x^{\alpha}\partial_{\alpha}-y_{\alpha}\partial^{\alpha}\\
P_{\mu}  & =\partial_{\mu}+\left(  y_{\mu}y_{\nu}-\tfrac{1}{2}y^{2}\eta
_{\mu\nu}\right)  \partial^{\nu}\\
K^{\mu}  & =\partial^{\mu}+\left(  x^{\mu}x^{\nu}-\tfrac{1}{2}x^{2}\eta
^{\mu\nu}\right)  \partial_{\nu}%
\end{align*}
with $\partial_{\mu}$ as before and $\partial^{\mu}=\tfrac{\partial}{\partial
y_{\mu}}$. These vector fields generate the following transformations:
\begin{align*}
&  \left.
\begin{array}
[c]{c}%
\tilde{x}^{\mu}=\Lambda_{\;\nu}^{\mu}x^{\nu}\\
\tilde{y}_{\mu}=y_{\nu}\Lambda_{\;\mu}^{\nu}\;
\end{array}
\right\}  \qquad\Lambda_{\;\nu}^{\mu}\in O(n-1,1)\\
&  \left.
\begin{array}
[c]{l}%
\tilde{x}^{\mu}=\lambda x^{\mu}\\
\tilde{y}_{\mu}=\lambda^{-1}y_{\mu}\;
\end{array}
\right\}  \qquad\lambda\in\mathbf{R}^{+}\\
&  \left.
\begin{array}
[c]{l}%
\tilde{x}^{\mu}=x^{\mu}+a^{\mu}\\
\tilde{y}_{\mu}=\dfrac{y_{\mu}-y^{2}\eta_{\mu\nu}a^{\nu}}{1-2y_{\nu}a^{\nu
}+a^{2}y^{2}}\;
\end{array}
\right\}  \qquad a^{\mu}\in\mathbf{R}^{n}\\
&  \left.
\begin{array}
[c]{l}%
\tilde{x}^{\mu}=\dfrac{x^{\mu}-x^{2}\eta^{\mu\nu}b_{\nu}}{1-2x^{\nu}b_{\nu
}+b^{2}x^{2}}\\
\tilde{y}_{\mu}=y_{\mu}+b_{\mu}\;
\end{array}
\right\}  \qquad b_{\mu}\in\mathbf{R}^{n}%
\end{align*}
We see that this construction treats $P_{\mu}$ and $K^{\mu}$ on an equal
footing: each generates translations in one sector and special conformal
transformations in the other. We retain the name \textit{translations} for the
transformations generated by $P_{\mu}$ and suggest the name
\textit{co-translations} for the ones corresponding to $K^{\mu}$. This action
does not mix the $x$- and $y$-sectors; angles are preserved in each of them
separately. If one sets either $x^{\mu}$ or $y_{\mu}$ to zero, the action
reduces to the standard action on compactified Minkowski space.

The isotropy subgroup at the origin $(x^{\mu},y_{\mu})=(0,0)$ is generated by
the vector fields $M_{\nu}^{\mu}$ and $D$; it is the similarity group $S(n)$,
so that indeed $B(n)=C(n)/W(n)$. We will refer to $B(n)$ as
\textit{homogeneous biconformal space}. In this geometry, symmetry of the
fibers is exchanged for increased coordinate freedom for the base manifold.
The tangent space at the origin of $B(n)$ is spanned by $P_{\mu}$ and $K^{\mu
}$. Notice that the quantity $x^{\mu}y_{\mu}$ is a fiber invariant: $D(x^{\mu
}y_{\mu})=0=M_{\beta}^{\alpha}(x^{\mu}y_{\mu})$.

As every homogeneous space, $B(n)$ may be considered the base space of a
principal bundle $C(n)\rightarrow B(n)$ with fiber $W(n)$. A connection on
this bundle follows from the canonical form on $C(n)$. Switching to
orthonormal indices $a,b...=1...n$, we write this form as%
\[
\mathbf{\omega}=M_{b}^{a}\mathbf{\omega}_{a}^{b}+P_{a}\mathbf{\omega}%
^{a}+K^{a}\mathbf{\omega}_{a}+D\mathbf{\omega}^{0},
\]
where the index positions on the basis $1$-forms $\{\mathbf{\omega}_{a}%
^{b},\mathbf{\omega}^{a},\mathbf{\omega}_{a},\mathbf{\omega}^{0}\}$ again
indicate the conformal weight. The form $\mathbf{\omega}$ satisfies the
Maurer-Cartan structure equations of the conformal group,
\begin{align}
\mathbf{d\omega}_{b}^{a}  & =\mathbf{\omega}_{b}^{c}\mathbf{\omega}_{c}%
^{a}-2\Delta_{cb}^{ad}\mathbf{\omega}_{d}\mathbf{\omega}^{c}\label{Str1}\\
\mathbf{d\omega}^{a}  & =\mathbf{\omega}^{b}\mathbf{\omega}_{b}^{a}%
+\mathbf{\omega}^{0}\mathbf{\omega}^{a}\label{Str2}\\
\mathbf{d\omega}_{a}  & =\mathbf{\omega}_{a}^{b}\mathbf{\omega}_{b}%
+\mathbf{\omega}_{a}\mathbf{\omega}^{0}\label{Str3}\\
\mathbf{d\omega}^{0}  & =-\mathbf{\omega}^{a}\mathbf{\omega}_{a},\label{Str4}%
\end{align}
where we leave off the wedges between adjacent forms. They are integrable by
virtue of the Jacobi identity. Clearly, the structure equations are invariant
under the involutive automorphisms (\ref{discreete}).

It is well known that for a symmetric space the $\frak{h}$-component of the
canonical form $\mathbf{\omega}$ defines a unique connection in the bundle
$G\rightarrow G/H$ called the canonical connection, with the following
properties: (a) It is invariant under both $G$ and the involutive automorphism
defining the symmetric space, (b) it has vanishing torsion, (c) it has
covariantly constant curvature, and (d) it is compatible with the
$G$-invariant indefinite Riemannian metric (\ref{Killing}). Thus the form
$\mathbf{\tilde{\omega}}=M_{b}^{a}\mathbf{\omega}_{a}^{b}+D\mathbf{\omega}%
^{0}$ defines the canonical connection on $C(n)\rightarrow B(n)$.

The cotangent space to $B(n)$ is spanned by the $2n$ differential forms
$\{\mathbf{\omega}^{a},\mathbf{\omega}_{a}\}$. Both $\mathbf{\omega}^{a}$ and
$\mathbf{\omega}_{a}$ can be consistently set to zero in the structure
equations (\ref{Str1})-(\ref{Str4}), which yields two classes of
$n$-dimensional subspaces of $B(n)$. We observe that the space $B(n)$ has a
natural symplectic structure: Since the connection forms $\mathbf{\omega}^{a}$
and $\mathbf{\omega}_{a}$ are independent, the $2$-form $\mathbf{\Omega
}:=\mathbf{\omega}^{a}\mathbf{\omega}_{a}$ is non-degenerate and, by virtue of
equation (\ref{Str4}), closed.

In addition to the symplectic structure $\mathbf{\Omega}$ and the conformally
invariant metric $K_{\alpha\beta}$, there exists the canonical almost complex
structure $J$ on homogeneous biconformal space It is integrable and compatible
with the symplectic structure in the sense that $\mathbf{\Omega(}%
Ju,Jv)=\mathbf{\Omega}(u,v)$, which implies that the $2n$-dimensional real
space under consideration is equivalent to an $n$-dimensional complex space
with K\"{a}hler metric $g(u,v)=\mathbf{\Omega}(Ju,v)$. Neither Minkowski
spaces, flat Weyl geometries, nor homogeneous conformal spaces give rise to
these structures in an intrinsic fashion.

Finally, we mention that a homogeneous biconformal space can serve as the
underlying model geometry for a curved generalization called
\textit{biconformal space}, which was introduced in \cite{Wheeler1}. We have
previously shown \cite{WW} that such generalized geometries permit linear
scale-invariant Lagrangians in any dimension and reproduce general relativity
on certain subspaces. They do not give rise to unphysical size changes.

In summary, we compare some of the properties of the homogeneous spaces
mentioned in this article:
\[%
\begin{tabular}
[c]{l|l|l|l|l}%
Homogeneous Space & Minkowski & Weyl & Conformal & Biconformal\\\hline
Group $G$ & Poincar\'{e} & $IS(n)$ & Conformal & Conformal\\
Subgroup $H$ & Lorentz & $S(n)$ & $IS(n)$ & $S(n)$\\
$dim(G/H)$ & $n$ & $n$ & $n$ & $2n$\\\hline
Killing form non-degenerate? & no & no & yes & yes\\
Symmetric space? & yes & yes & no & yes\\
Local scale-invariance? & no & yes & yes & yes\\
Symplectic structure? & no & no & no & yes\\
K\"{a}hler structure? & no & no & no & yes
\end{tabular}
\]
\pagebreak

\end{document}